\documentclass[twocolumn,showpacs,superscriptaddress,preprintnumbers,
amsmath,amssymb]{revtex4}
\usepackage{graphicx}
\usepackage{dcolumn}
\usepackage{bm}

\newcommand{\mean}[1]{\left\langle #1 \right\rangle}

\def\simge{\mathrel{
     \rlap{\raise 0.511ex \hbox{$>$}}{\lower 0.511ex \hbox{$\sim$}}}}
\def\simle{\mathrel{
     \rlap{\raise 0.511ex \hbox{$<$}}{\lower 0.511ex \hbox{$\sim$}}}}

\begin{document}
\preprint{SPhT-T06/077; TIFR/TH/06-20}
\title{Eccentricity fluctuations and elliptic flow at RHIC}
\author{Rajeev S. Bhalerao}
\affiliation{Department of Theoretical Physics, TIFR,
   Homi Bhabha Road, Colaba, Mumbai 400 005, India}
\author{Jean-Yves Ollitrault}
\affiliation{Service de Physique Th\'eorique, CEA/DSM/SPhT, Unit\'e de
recherche associ\'ee au CNRS,\\ F-91191 Gif-sur-Yvette Cedex, France.}

\date{\today}
\begin{abstract}
Fluctuations in nucleon positions can affect the spatial 
eccentricity of the overlap zone in nucleus-nucleus collisions. 
We show that elliptic flow should be scaled by different 
eccentricities depending on which method is used for the 
flow analysis.
These eccentricities are estimated semi-analytically.
When $v_2$ is analyzed from 4-particle cumulants, or using the 
event plane from directed flow in a zero-degree calorimeter, 
the result is shown to be insensitive to eccentricity fluctuations. 
\end{abstract}

\pacs{25.75.Ld, 24.10.Nz}

\maketitle

\medskip
\noindent 1. Introduction

\begin{figure}
\includegraphics*[width=0.7\linewidth]{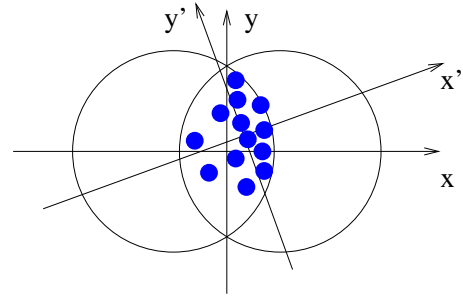}
\caption{Schematic view of a collision of two identical
nuclei, in the plane transverse to the beam direction ($z$-axis). 
The $x$- and $y$-axes are drawn as per the standard convention. 
The dots indicate the positions of 
participant nucleons. Due to fluctuations, the 
overlap zone could be shifted and tilted with respect to the $(x,y)$
frame. $x'$ and $y'$ are the principal axes of inertia of the dots.}
\label{fig:ellipse}
\end{figure}

Elliptic flow, $v_2$, is one of the key observables in 
nucleus-nucleus collisions at RHIC~\cite{Ackermann:2000tr}. 
It originates from the almond shape of the overlap zone 
(see Fig.~\ref{fig:ellipse}) which produces, through unequal 
pressure gradients, an anisotropy in the transverse momentum 
distribution~\cite{Ollitrault:1992bk}, the so-called 
$v_2\equiv\mean{\cos 2\phi}$, where $\phi$'s are the azimuthal
angles of the detected particles with respect to the
reaction plane. 

Preliminary analyses of $v_2$ in Cu-Cu collisions at RHIC
~\cite{Masui:2005aa,Manly:2005zy,Wang:2005ab},
presented at the QM'2005 conference, 
reported values surprisingly large compared to 
theoretical expectations,
almost as large as in Au-Au collisions. 
It was shown by the PHOBOS collaboration~\cite{Manly:2005zy}
that fluctuations in nucleon positions provide a natural explanation 
for this large magnitude. 
The idea is the following:
The time scale of the nucleus-nucleus collision at RHIC is so 
short that each nucleus sees the nucleus coming in the opposite 
direction in a frozen configuration, with nucleons located at positions 
whose probabilities are determined according to the nuclear wave
function. 
Fluctuations in the nucleon positions result in fluctuations 
in the almond shape and orientation (see Fig.~\ref{fig:ellipse}), 
and hence in larger values of $v_2$. 

In this Letter, we discuss various definitions of the 
eccentricity of the overlap zone. 
We show that estimates of $v_2$ using different 
methods should be scaled by appropriate choices of the 
eccentricity.
We then compute the effect of fluctuations 
on the eccentricity semi-analytically to leading order in $1/N$, 
where $N$ is the mean number of participants at a given centrality.
A similar study was recently performed by S. Voloshin 
on the basis of Monte-Carlo Glauber calculations~\cite{Voloshin:2006gz}. 

\medskip
\noindent 2. Eccentricity scaling and fluctuations

Elliptic flow 
is determined by the initial density profile. 
Although its precise value depends on the detailed shape of the
profile, most of the relevant information is encoded in 
three quantities: 
1) the initial 
eccentricity of the overlap zone, $\varepsilon$, which will be defined 
more precisely below;
2) the density $n$, which determines 
pressure gradients through the equation of state 
(by density, we mean the particle density, $n$, 
at the time when elliptic flow develops; this time is of 
the order of the transverse size $R$. Quite remarkably, 
the density thus defined varies little with centrality, 
and has almost the same value in Au-Au and Cu-Cu collisions 
at the same colliding energy per nucleon~\cite{Bhalerao:2005mm});
3) the system transverse size $R$, which determines the number 
of collisions per particle. $v_2$ scales like $\varepsilon$ for 
small $\varepsilon$, that is, $v_2=\varepsilon f(n, R)$. 

This proportionality relation is only approximate. However, 
hydrodynamical calculations~\cite{Bhalerao:2005mm} show that it is a
very good approximation in practice for nucleus-nucleus collisions. 
Eccentricity scaling holds for integrated flow 
as well as for the differential flow of identified particles. 
In the latter case, the function $f(n,R)$ also depends on 
the mass, transverse momentum and rapidity of the particle.

Eccentricity scaling of $v_2$ is generally believed to be
a specific prediction of relativistic 
hydrodynamics. In the form above, the scaling is expected 
to be more general: it does not require 
thermalization, as implicitly assumed by hydrodynamics. 
If thermalization is achieved, that is, if the system size $R$ 
is much larger than the mean free path $\lambda$, then the scaling
is stronger: $v_2/\varepsilon$ no longer depends on $R$, 
but only on the density $n$~\cite{Bhalerao:2005mm}. 

The standard definition of the eccentricity is~\cite{Sorge:1998mk} 
\begin{equation} 
\label{epsstandard}
\varepsilon_{\rm s}=
\frac{\mean{y^2-x^2}} {\mean{y^2+x^2}},
\end{equation}
where $(x,y)$ is the position of a participant nucleon in 
the coordinate system defined in Fig.~\ref{fig:ellipse}.
Throughout this paper, $\mean{...}$ denotes an ensemble average: 
here, it means an average over participant nucleons and 
over many collision events of the same impact parameter. 
This standard eccentricity applies to most hydrodynamic 
calculations: indeed, most hydrodynamic calculations 
(with the exception of Ref.~\cite{Socolowski:2004hw}) use smooth,  
event-averaged initial conditions, with an initial entropy density 
proportional to the density of participants~\cite{Kolb:2003dz}. 
The charged multiplicity then scales like the number of 
participants, as observed experimentally. 
(Adding a component proportional to the number of 
binary collisions, as argued in~\cite{Kharzeev:2000ph}, does not
change the eccentricity significantly). 

It was recently argued~\cite{Drescher:2006pi}
 that the Color Glass Condensate picture of heavy-ion 
collision leads to a different definition of the eccentricity, which
may be significantly larger than the standard eccentricity. This 
interesting possibility will not be considered further here. 

Now, because of the event-by-event fluctuations in the participant 
nucleon positions~\cite{Miller:2003kd}, the eccentricity driving 
elliptic flow in a given event is that defined by the principal 
axes $(x',y')$ of the distribution of participant nucleons, 
see Fig.~\ref{fig:ellipse}. 
This ``participant eccentricity'' $\varepsilon_{\rm part}$ 
can be written as~\cite{Manly:2005zy}
\begin{equation}
\label{epspart}
\varepsilon_{\rm part}=\frac{
\sqrt{(\sigma_y^2-\sigma_x^2)^2+4 \sigma_{xy}^2}}
{\sigma_y^2+\sigma_x^2},
\end{equation}
where 
\begin{eqnarray}
\label{defsigma}
\sigma_x^2&=&\{x^2\}-\{x\}^2\cr
\sigma_y^2&=&\{y^2\}-\{y\}^2\cr
\sigma_{xy}&=&\{xy\}-\{x\}\{y\},
\end{eqnarray}
and $\{...\}$ denotes the average over all participants in one 
collision event (sample average). 
Our basic assumption in this paper is that {\it in each event\/},
the elliptic flow $v_2$ is proportional to the participant
eccentricity $\varepsilon_{\rm part}$. 

Experimentally, elliptic flow is analyzed by selecting events in a
centrality class. The quantities which drive elliptic flow, 
namely, the density $n$, the transverse size $R$, the eccentricity 
$\varepsilon_{\rm part}$, fluctuate from one event to the other.
This causes dynamical fluctuations of $v_2$. 
The effect of impact parameter fluctuations was  carefully 
studied in~\cite{Adler:2002pu} and was shown to be small. 
In this paper, we focus on fluctuations in the positions of 
participant nucleons. 
It will be shown below that fluctuations in the eccentricity 
$\varepsilon_{\rm part}$ dominate over fluctuations in size and density.

\medskip
\noindent 3. To each method its own eccentricity

Since the reaction plane is not known exactly on an event-by-event basis, 
$v_2$ is measured indirectly using azimuthal correlations. 
Several methods have been used, which yield different estimates
of $v_2$. 
We argue that these estimates are affected in different ways
by fluctuations of $v_2$:
\begin{itemize}
\item The event-plane method~\cite{Poskanzer:1998yz} has been 
implemented by the STAR~\cite{Ackermann:2000tr}, 
PHOBOS~\cite{Back:2002gz} and PHENIX~\cite{Adler:2003kt} 
collaborations at RHIC. The event plane is an estimate of the 
reaction plane; it is defined as the plane spanned by the collision 
axis and the major axis of the ellipse formed by the transverse momenta of 
outgoing particles. 
It corresponds to the $x'$ axis defined by the participants 
(Fig.~\ref{fig:ellipse}), up to statistical
fluctuations which are taken care of by the analysis 
(this is the so-called ``event-plane resolution''). 
Outgoing particles are then individually correlated to this event
plane. The corresponding estimate of $v_2$ will be denoted by 
$v_2\{ {\rm EP2}\}$. 
Another method determines $v_2$ from two-particle azimuthal 
correlations between outgoing particles~\cite{Adcox:2002ms}. 
The corresponding estimate is usually denoted by $v_2\{2\}$, and 
is defined by an equation of the type 
$v_2\{2\}=\sqrt{\mean{\cos 2(\phi_1-\phi_2)}}$. 
Both methods are essentially equivalent~(see Sec.~III.C.2 of 
Ref.~\cite{Borghini:2000sa}). 
If $v_2$ fluctuates, both yield the rms value of 
$v_2$~\cite{Adler:2002pu}: 
$v_2\{2\}\simeq v_2\{{\rm EP2}\}=\sqrt{\mean{v_2^2}}$. 
Since $v_2$ in each event scales with 
$\varepsilon_{\rm part}$, one expects 
$v_2\{2\}\simeq v_2\{{\rm EP2}\}\propto\varepsilon\{2\}$, where 
$\varepsilon\{2\}$ is defined by~\cite{Miller:2003kd}
\begin{equation}
\label{v2std}
\varepsilon\{2\}\equiv\sqrt{\mean{\varepsilon_{\rm
    part}^2}}. 
\end{equation}
This scaling differs from that proposed in Ref.~\cite{Manly:2005zy}, 
$v_2\propto\mean{\varepsilon_{\rm part}}$. 
\item Four-particle cumulants of azimuthal 
correlations~\cite{Borghini:2001vi} can be used to 
reduce the bias from nonflow correlations. The corresponding 
estimate of $v_2$ is denoted by $v_2\{4\}$. It involves a combination
of 2-particle and 4-particle correlations, 
i.e., the 2nd and 4th moments of the distribution of 
$v_2$, $\mean{v_2^2}$ and $\mean{v_2^4}$. Assuming again that 
$v_2$ scales with $\varepsilon_{\rm part}$ in each event, 
one obtains $v_2\{4\}\propto\varepsilon\{4\}$, where 
$\varepsilon\{4\}$ is defined by~\cite{Miller:2003kd}: 
\begin{equation}
\label{v24}
\varepsilon\{4\}\equiv\left(2\mean{\varepsilon_{\rm part}^2}^2
-\mean{\varepsilon_{\rm part}^4}\right)^{1/4}.
\end{equation}

\item An alternative method determines the event-plane from directed 
flow~\cite{Adams:2003zg}, determined in a zero-degree calorimeter
(ZDC) which detects spectator neutrons. Particles in the central
rapidity region are then correlated to this event plane. This 
last estimate of $v_2$ is denoted by 
$v_2\{ {\rm ZDC}\}$~\cite{Wang:2005ab}. 
It differs from the previous ones in that the reference direction
(the event plane) is determined by spectator neutrons from the 
projectile, rather than by participants. The direction defined by 
spectator neutrons is the $x$-axis, not the $x'$-axis 
(see Fig.~\ref{fig:ellipse}), up to fluctuations in spectator 
positions which are taken care of by the analysis. 
Therefore, the relevant eccentricity for this analysis 
is the reaction-plane eccentricity
\begin{equation}
\varepsilon_{\rm RP}\equiv \frac{\sigma_y^2-\sigma_x^2}
{\sigma_y^2+\sigma_x^2},
\end{equation}
and $v_2\{ {\rm ZDC} \}$ should scale correspondingly like 
\begin{equation}
\label{v2ZDC}
v_2\{ {\rm ZDC}\}\propto \mean{\varepsilon_{\rm RP}}. 
\end{equation}
\end{itemize}
This estimate of $v_2$ is also unbiased by nonflow correlations 
because it involves a 3-particle correlation (instead of 
a 2-particle correlation in the standard event-plane method), 
and also because the ZDC calorimeter has a wide rapidity gap 
with the central rapidity detector. 

\medskip
\noindent 4. Computing the fluctuations

We now compute the eccentricities entering Eqs.~(\ref{v2std}), 
(\ref{v24}) and (\ref{v2ZDC}), namely, $\varepsilon\{2\}$, 
$\varepsilon\{4\}$ and $\mean{\varepsilon_{\rm RP}}$, to 
leading order in the fluctuations.

To that end, we write each sample average (over one event) 
entering Eq.~(\ref{defsigma}) as the sum of an ensemble average and 
a fluctuation, e.g., $\{x^2\}=\mean{x^2}+~\delta_{x^2}$, 
where $\delta_{x^2}$ is the fluctuation. 
We choose the coordinate system in Fig.~\ref{fig:ellipse}, where
$\mean{x}=\mean{y}=\mean{xy}=0$. 
Substituting in Eq.~(\ref{epspart}) and retaining terms 
to second order in the $\delta$'s gives
\begin{eqnarray}
\label{epspart2}
\varepsilon_{\rm part}=\varepsilon_{\rm s}
&+&\frac{\delta_{y^2}-\delta_{x^2}}{\mean{r^2}}
-\varepsilon_{\rm s}\frac{\delta_{y^2}+\delta_{x^2}}{\mean{r^2}}
-\frac{\delta_y^2-\delta_x^2}{\mean{r^2}} \nonumber  \\
&+&\frac{2\delta_{xy}^2}{\varepsilon_{\rm s}\mean{r^2}^2}
-\frac{\delta_{y^2}^2-\delta_{x^2}^2}{\mean{r^2}^2} \nonumber  \\
&+&\varepsilon_{\rm s}\frac{\delta_y^2+\delta_x^2}{\mean{r^2}}
+\varepsilon_{\rm s}\frac{(\delta_{y^2}+\delta_{x^2})^2}{\mean{r^2}^2},
\end{eqnarray}
where $\mean{r^2}=\mean{x^2+y^2}$. 
The reaction plane eccentricity $\varepsilon_{\rm RP}$ is given
by a similar expression, except for the fifth term on the right-hand
side (rhs) proportional to  $\delta_{xy}^2$ which does not appear in 
$\varepsilon_{\rm RP}$. 
The 2nd and 3rd terms on the rhs
of Eq.~(\ref{epspart2}) are linear and the remaining 5 terms are quadratic in
fluctuations. In an obvious notation Eq.~(\ref{epspart2}) can be rewritten as 
\begin{equation}
\label{AandB}
\varepsilon_{\rm part}=\varepsilon_{\rm s} +\sum_{i=1}^2 A_i\delta_i^{(1)}
+\sum_{i=1}^5 B_i\delta_i^{(2)}, 
\end{equation}
where $\delta_i^{(1)}$ are linear and $\delta_i^{(2)}$ are quadratic
in $\delta$'s. $\varepsilon_{\rm RP}$ is given by a similar
expression, with one less quadratic term. 
It is now straightforward to derive the expressions for
$\mean{\varepsilon_{\rm part}}, \mean{\varepsilon^2_{\rm part}}, 
\mean{\varepsilon^4_{\rm part}}$, etc. We get
\begin{eqnarray}
\label{epsmoments}
\mean{\varepsilon_{\rm part}}&=& \varepsilon_{\rm s} 
+ \mean{B_i\delta_i^{(2)}}\nonumber \\
\mean{\varepsilon_{\rm part}^2}&=& \varepsilon_{\rm s}^2
+\mean{(A_i\delta_i^{(1)})^2}+
2\varepsilon_{\rm s}\mean{B_i\delta_i^{(2)}}\nonumber \\
\mean{\varepsilon_{\rm part}^4}&=& \varepsilon_{\rm s}^4
+6 \varepsilon_{\rm s}^2 \mean{(A_i\delta_i^{(1)})^2}
+4 \varepsilon_{\rm s}^3 \mean{B_i\delta_i^{(2)}}, 
\nonumber \\
\end{eqnarray}
where we have retained terms to second order in the $\delta$'s and used
$\mean{\delta_i^{(1)}}=0$; summation over the repeated index is
understood. 

The averages on the rhs of Eqs.~(\ref{epsmoments}) 
are easily computed by using the following identity, which holds
for $N$ independent participant nucleons 
\begin{equation}
\mean{\delta_f\delta_g} =
\frac{\mean{fg}-\mean{f}\mean{g}}{N},
\end{equation}
where $f$ and $g$ are any functions of $x$ and $y$. We get 
\begin{eqnarray}
\label{epscumul2}
\varepsilon\{2\}^2
&=&\varepsilon_{\rm s}^2+\cr
&&\frac{\mean{r^4}}{N\mean{r^2}^2}
\left(1+3 \varepsilon_{\rm s}^2+
4\varepsilon_{\rm s}\frac{\mean{r^4 \cos 2\phi}}{\mean{r^4}}\right),
\end{eqnarray}
\begin{equation}
\label{epscumul4}
\varepsilon\{4\}^4
=\varepsilon_{\rm s}^4+
\frac{2\mean{r^4}}{N\mean{r^2}^2}
\left(\varepsilon_{\rm s}^4-
\varepsilon_{\rm s}^2\frac{\mean{r^4 \cos  4\phi}}{\mean{r^4}}\right),
\end{equation}
and 
\begin{equation}
\label{epsRP}
\mean{\varepsilon_{\rm RP}}=
\varepsilon_{\rm s}
+\frac{\mean{r^4}}{N\mean{r^2}^2}\left(\varepsilon_{\rm s}+
\frac{\mean{r^4 \cos 2\phi}}{\mean{r^4}}\right),
\end{equation}
where $(r,\phi)$ are the polar coordinates in the $(x,y)$ plane.

\begin{figure}
\includegraphics*[width=\linewidth]{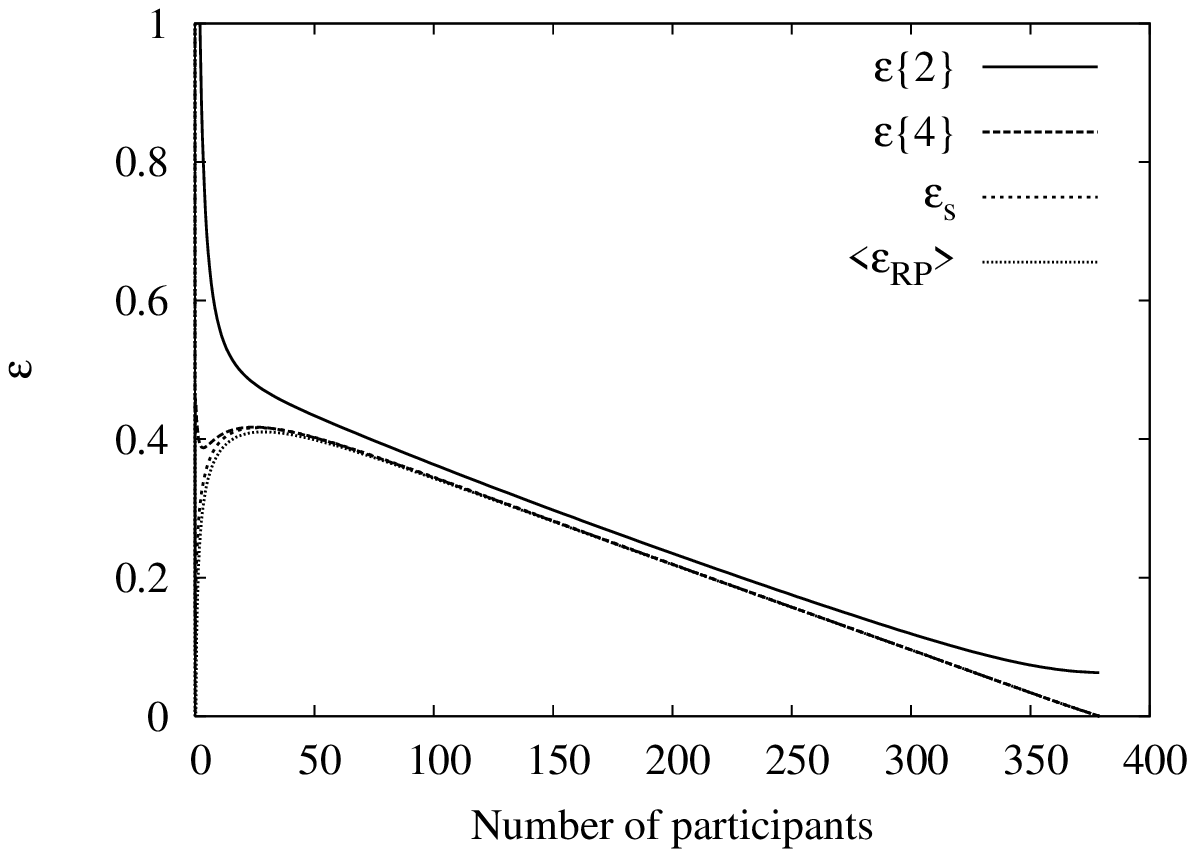}
\includegraphics*[width=\linewidth]{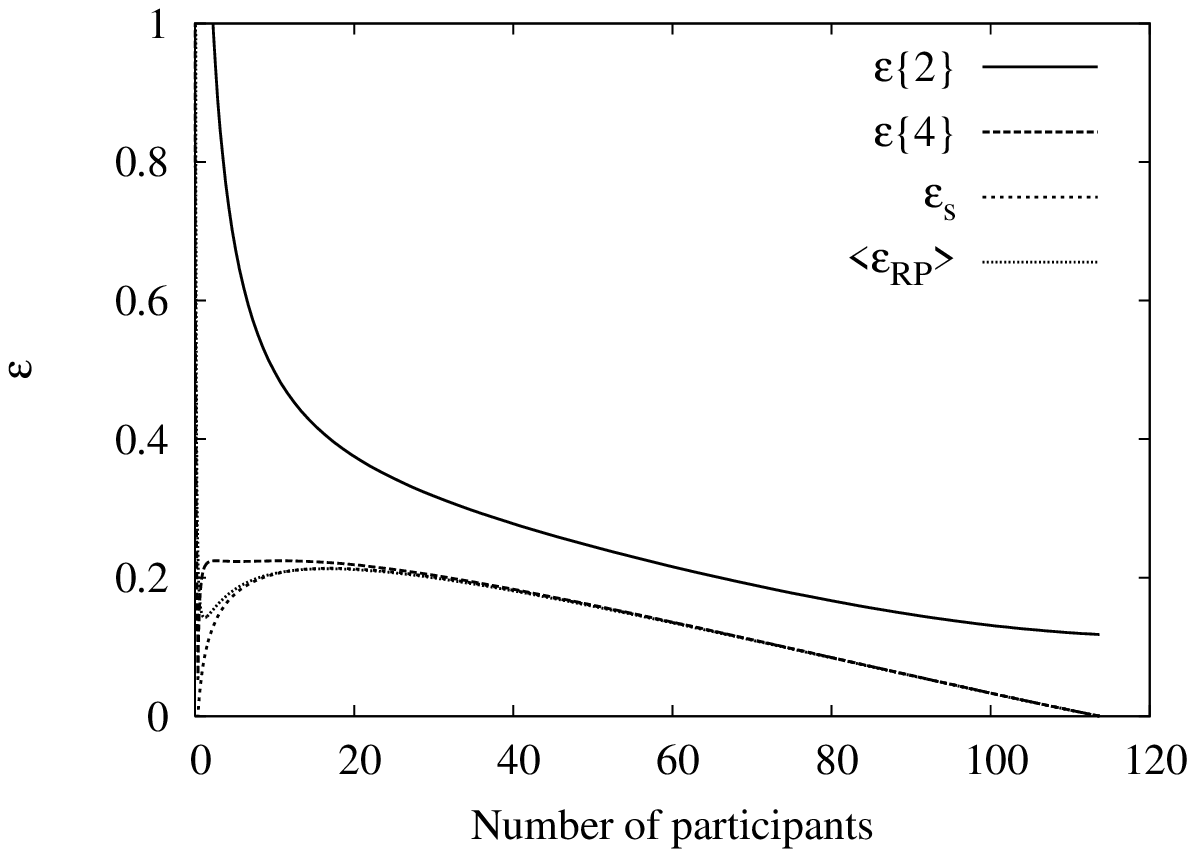}
\caption{Standard eccentricity $\varepsilon_{\rm s}$,
participant eccentricities 
$\varepsilon\{2\}$ and $\varepsilon\{4\}$, and reaction-plane 
eccentricity $\mean{\varepsilon_{\rm RP}}$, 
vs the number of participant nucleons for a Au-Au collision (top)
and a Cu-Cu collision (bottom).}
\label{fig:Au_Npart}
\end{figure}

The number of participants, $N$, and the moments of their distribution, 
$\mean{r^4}$, $\mean{r^4\cos2\phi}$, \dots, 
are easily computed numerically in the Glauber model with Woods-Saxon
distribution of participants. 
The parameters of our Glauber calculation 
are the same as in Ref.~\cite{Kharzeev:2000ph}, with one minor 
difference: 
we neglect fluctuations in the number of participants, i.e., we assume
a one-to-one relation between impact parameter and number of
participants. 
Orders of magnitude of the various coefficients involved are 
$\mean{r^4}/\mean{r^2}^2\sim 1.4-2$, 
$\mean{r^4\cos 2\phi}/\mean{r^4}\sim-\varepsilon_{\rm s}$, 
$\mean{r^4\cos 4\phi}/\mean{r^4}\sim \varepsilon_{\rm s}^2$. 
Numerical results for the various eccentricities as a function of 
the number of participants are given in Fig.~\ref{fig:Au_Npart} 
for Au-Au and Cu-Cu collisions.

The standard eccentricity $\varepsilon_{\rm s}$ vanishes for the most
central as well as for the most peripheral collisions as expected. 
On the other hand, $\varepsilon\{2\}$ rises steeply as 
$N$ decreases; this is due to the $1/N$ contribution of the 
fluctuations, see Eq.~(\ref{epscumul2}). 
This contribution originates from the 
$\delta_{y^2}-\delta_{x^2}$ and $\delta_{xy}^2$ terms in 
Eq.~(\ref{epspart2}).
The number of participants $N$ being smaller for the Cu-Cu collision than
for the Au-Au collision, the magnitude of the fluctuations is
relatively larger in the former case.

The 4-cumulant $\varepsilon\{4\}$ is only slightly larger than 
$\varepsilon_{\rm s}$, down to low values of $N$.
This is because the $1/N$ term in Eq.~(\ref{epscumul4}) 
is multiplied by a term of order $\varepsilon_{\rm s}^4$,
As a result, the {\it relative\/} difference between 
$\varepsilon\{4\}$ and $\varepsilon_{\rm s}$ is only of order
$1/N$. Similarly, the reaction plane eccentricity, 
$\mean{\varepsilon_{\rm RP}}$, is almost equal to 
$\varepsilon_{\rm s}$. 

We have considered so far the fluctuations in the eccentricity of the
overlap zone. We now briefly discuss fluctuations in the size or the
transverse area of the overlap zone, $S\sim 2\pi\sigma_x\sigma_y$.
They can be calculated exactly in the same way as fluctuations in 
the eccentricity. The result is 
\begin{equation}
\frac{\mean{S^2}-\mean{S}^2}{\mean{S}^2}=\frac{1}{4N}
\left(\frac{\mean{y^4}}{\mean{y^2}^2}+
\frac{\mean{x^4}}{\mean{x^2}^2}+
\frac{2\mean{x^2y^2}}{\mean{x^2}\mean{y^2}}-4\right).
\end{equation}
Numerically, these fluctuations are found to be practically
negligible. The reason why eccentricity fluctuations are important 
is that the $1/N$ term in Eq.~(\ref{epscumul2}) must be compared with 
$\varepsilon_{\rm s}^2$, which is itself a small number.

\medskip
\noindent 5. Discussion

We have shown that the values of $v_2$ analyzed with different methods 
should be scaled by different eccentricities:
$v_2\{ {\rm EP2}\}$ (standard $v_2$), 
$v_2\{4\}$ and $v_2\{ {\rm ZDC}\}$ 
should be scaled respectively by $\varepsilon\{2\}$,
$\varepsilon\{4\}$ and $\mean{\varepsilon_{\rm RP}}$, 
defined in Eqs.~(\ref{v2std}), (\ref{v24}) and (\ref{v2ZDC}). 

Our most important result is that $\varepsilon\{4\}$ and 
$\mean{\varepsilon_{\rm RP}}$ are almost equal to the 
standard eccentricity, while $\varepsilon\{2\}$ is strongly
affected by fluctuations for small systems and/or peripheral
collisions. 
An important contribution to the fluctuations comes from 
the angle between the $x'$-axis and the 
$x$-axis in Fig.~\ref{fig:ellipse}, i.e., from the angle of tilt 
of the participant ellipse relative to the reaction plane. 
This effect was neglected in Ref.~\cite{Miller:2003kd} and 
first taken into account in Ref.~\cite{Manly:2005zy}. 
The ZDC analysis eliminates the effect of the tilt angle 
by measuring the eccentricity in the (x,y) axes; in the case of
the 4-cumulant analysis, most of the fluctuations happen to 
cancel in the subtraction of Eq.~(\ref{v24}).

Our results show that higher-order estimates 
of elliptic flow, $v_2\{4\}$ and $v_2\{\rm ZDC\}$, are not only 
insensitive to nonflow effects, but also, to a large extent, 
to fluctuations in the participant eccentricity.
This is confirmed by transport calculations~\cite{Zhu:2005qa}, and 
explains the observed agreement between $v_2\{4\}$ and 
$v_2\{\rm ZDC\}$ in Au-Au collisions~\cite{Wang:2005ab}. 

Fluctuations in the participant eccentricity, on the other hand,  
tend to increase the value of the standard, event-plane $v_2$. 
We have estimated this increase quantitatively, assuming 
independent nucleons. 
Within this simple model, fluctuations  account for less than 
one half of the observed difference between $v_2\{2\}$ 
and $v_2\{4\}$ in Au-Au collisions~\cite{Wang:2005ab}. 
The remaining difference can be (at least partly) 
ascribed to nonflow effects. Nonflow effects are 
clearly seen in the different $p_T$-dependences of $v_2\{2\}$ and 
$v_2\{4\}$, which cannot be explained by fluctuations. 

However, one should be aware that estimates of the 
participant eccentricity are model dependent:
in particular, Monte-Carlo Glauber 
calculations~\cite{Manly:2005zy,Voloshin:2006gz} generally 
yield higher fluctuations. 
The difference is the following: in a Monte-Carlo Glauber, 
each nucleon is modelled as black disk of transverse area $\sigma$, 
and a nucleon from nucleus A is a participant if it overlaps 
with at least a nucleon from nucleus B, and vice-versa.
Participant nucleons are therefore correlated, and the black-disk
approximation maximizes these correlations. 
Due to such correlations, which are neglected in our
calculation, our estimate of the participant eccentricity can be 
considered a lower bound. 
Quantitatively, correlations can increase the effect of fluctuations
by a factor up to 2: if the participant nucleons can only be found 
in pairs of overlapping disks, this amounts to replacing $N$ by 
$N/2$ (the number of pairs) in Eqs.~(\ref{epscumul2}-\ref{epsRP}). 

Finally, let us compare our results with the recent Monte-Carlo 
Glauber calculations of Ref.~\cite{Voloshin:2006gz}. 
The results are in qualitative agreement with ours: 
$\varepsilon\{4\}$ is much closer to $\varepsilon_{\rm s}$ 
than to $\varepsilon\{2\}$ for moderate centralities. 
For large impact parameters, however, $\varepsilon\{4\}$ is almost 
equal to $\varepsilon\{2\}$. This means that the participant 
eccentricity, although much larger than the standard eccentricity, 
fluctuates little from one event to the other. This intriguing
behaviour could be a consequence of the strong correlations 
mentioned above, and deserves further investigation. 

To summarize, the elliptic flow scaled by the eccentricity of the
overlap zone, $v_2/\varepsilon$, is an important observable at RHIC as
well as LHC, because if it is found to be independent of the system
size, one has a strong pointer toward thermalization. We have 
discussed various definitions of $v_2$ and $\varepsilon$ and studied how
they are affected by fluctuations in nucleon positions.
We have shown that when $v_2$ is analyzed using 4-particle cumulants or 
the event-plane from directed flow in a ZDC calorimeter, the resulting 
estimate essentially scales with the standard eccentricity, and is 
insensitive to the fluctuations in the participant eccentricity 
considered by PHOBOS~\cite{Manly:2005zy}.

\section*{Acknowledgments}

We acknowledge the financial support from CEFIPRA, New Delhi, under
its project no. 3104-3. JYO thanks G. Wang, G. Roland, M. Nardi, 
S. Manly and A. Poskanzer for discussions.

\end{document}